**Evaluation of the energy states of hydrogen atom by using Schrödinger equation with a Coulomb potential which is modified due to the interaction between the magnetic moments of the proton and of the electron**

**Voicu Dolocan**

*Faculty of Physics, University of Bucharest, Bucharest, Romania*

## ABSTRACT

By using a Coulomb potential, modified by the interaction between the magnetic moments of the electron and proton, we have calculatedthe energy levels of the hydrogen atom. We have obtained fine and hyperfine structure as well as the Lamb shift. All these effects are obtained from a simple formula which is a direct solution of the Schrödinger equation. The obtained results are in a good agreement with experimental data. For example, the hyperfine splitting between the energy levels of the states $1S_{1/2,1}$ and $1S_{1/2,}$o is of the order of $5.6\times10^{-6}$ eV, which is the source of the famous "21 cm line" which is strongly useful to radio astronomers for tracking hydrogen in the interstellar medium of galaxies. The energy of the states $nP_{1/2}$ is lower than those of the states $nS_{1/2}$ (Lamb shift) because in the first case the interaction between the magnetic moments of the proton and the electron spins is diminished by the spin-orbit coupling..

**Keywords:** Magnetic moments; fine and hyperfine structure; Lamb shift of hydrogen atom.

# 1. INTRODUCTION

With the usual Hamiltonian of the hydrogen-like atom we have the $n^2$-fold degeneracy states with the same principal quantum number, or $2n^2$ –fold once the spin degree of freedom is included. I this real world however, the degeneracy is lifted by corrections that arise due to the special relativity. These corrections (known as fine structure) derive from three (superficially) different sources: (*a*) relativistic corrections to the kinetic energy , (*b*) coupling between the spin and orbital degree of freedom, (*c*) and the contribution knowing as a Darwin term. Relativistic corrections split degenerate multiplets leading to small shift in energy, ca $10^{-4}$ – $10^{-6}$ eV. In additoon, nucleus has a spin which leads to a nuclear magnetic moment . Interaction of electronic magnetic moment with filed generated by nuclear magnetic moment leads to further splitting of multiplets ( hyperfine structure), ca $10^{-7}$ – $10^{-8}$ eV. In 1947, an experimental study by W. Lamb discovered that $2P_{1/2}$ state is slightly lower than $2S_{1/2}$ state – Lamb shift [1]. The effect is expalined by the theory of quantul electrodynamics [2], in which the electromagnetic interaction itself is quantized. Some of the effects of this theory which cause the Lamb shift are as follows: vacuum polarization, electron mass renormalization, anomalous magnetic moment. On the basis of this theory we have studied in a previous paper [3] the Lamb shift without taking into account the electron charge. Famous fine structure was first gotten by Bohr-Sommerfeld model in 1816 [4]. The fine structure used formally now is the hydrogen solution by Dirac equation [5]. Surprisingly, these solutions by Dirac equations are just equal to those of Sommerfeld model. However, Dirac's hydrogen includes a lot of wrong states (= $1P_{1/2}$, $2D_{3/2}$, $3F_{5/2}$, …). The interpretation of very tiny Lamb shift depends completely on the interpretation that Dirac's hydrogen is right. Quantum electrodynamics Lamb shift is much more complicated and filled with artificial tricks. Lamb shift measurements is too difficult and vague in respect of accueacy. We cannot see what is really happening in the key small effect = 0.000004372 eV, 10n8 MHz) hyperfine level. Though the Lamb shift is very small, the author tried to measure this value believing $2S_{1/2}$ state is "metastable" and the collision between excited hydrogen atom and plates is a precise method for Lamb shift. In this experiment there is no guarantee that modified Zeeman effect is always linearly effective, and excited metastable statesreally means $2S_{1/2}$. There are only assumpions. And, of course, the collision method is rough and not precise to measure this very tiny value. Even the latest optic methods, cannot confirm these states really express the ebergy difference between $2S_{1/2}$ and $2P_{1/2}$. They just estimate it. Considering Lamb shift is almost same as nuclear hyperfine structure some nuclear or electron's vibrations may influence very tiny data. In this paper we calculate the hydrogen energy levels by solving Schrödinger equation with the modified Coulomb potential by interaction between the magnetic moments of nucleus and electron's respectively, as we have proceed to study ferromagnetism [6]. Also, we have used this modified Coulomb potential to evaluate high excitation energy levels of helium atom [7], deuteron energy states [8], and energy levels of a pionic atom[9]. As we will see below, Lamb shift appears as a natural result for the energy eigenvalues of Schrödinger equation.

## 2. EFFECTS OF THE INTERACTION BETWEEN THE MAGNETIC MOMENTS ON THE COULOMB POTENTIAL

In a previous paper [10] we have found the following expression for the energy of interaction between two electrons via bosons

$$E_i = -\frac{\hbar^3 D^2}{32 m^2 R^2 \left(\rho_o + \frac{DR}{c^2}\right)^2} \sum_{\mathbf{k},\mathbf{q},\mathbf{q_o}} \frac{(\mathbf{q}.\mathbf{q_o})^2}{\omega_{\mathbf{q}}^2 \omega_{\mathbf{q_o}}^2} \times \frac{1}{2} \left| \sum_n e^{i\mathbf{q_o}.\mathbf{R_n}} \right|^2 \frac{1}{\varepsilon_{\mathbf{k}} - \varepsilon_{\mathbf{k-q}} - \omega_{\mathbf{q}}} n_{\mathbf{k}} n_{\mathbf{k-q}} (n_q + 1)(n_{\mathbf{q_o}} + 1) \tag{1}$$

whbere $D$ is a coupling constant, $m$ is the mass of an electron, $R$ is the distance between the two electrons, $\rho_o$ is the massive density of the interacting field, $DR/c^2$ is the "mass density" associated to the energy of the ineracting field when this is not a massive field, $\omega_{\mathbf{q}}$ =cq is the classical oscillation frequency of the interacting field, $\omega_{\mathbf{q}o}$ is the oscillation frequency of an electron, $\mathbf{q}$ is the wave vector of the interacting field, $\mathbf{q}_o$ is the wave vector of the boson associated with the electron, $\mathbf{k}$ is the wave vector of the electron, $\varepsilon_{\mathbf{k}} = \hbar k^2/2m$, $n_{\mathbf{q}}$ is the occupation number of the bosons associated with the interacting field, $n_{\mathbf{qo}}$ is the occupation number of the bosons associated with an electron and $n_{\mathbf{k}}$ is the occupation number of the electrons. When the interacting filed is a photon field, then $\rho_o$ = 0. For a quasifree electron $\varepsilon_{\mathbf{k}} - \varepsilon_{\mathbf{k-q}} = 0$, $\omega_{\mathbf{qo}} = \hbar q_o^2/2m$. The Coulomb interaction occurs via photons, so that we may assume that the interacting electron oscillates with $\omega_{\mathbf{q}o}$. By using that $n_{\mathbf{q}} = n_{\mathbf{q}o} = 0$, $n_{\mathbf{k}} = n_{\mathbf{k}\text{-}\mathbf{q}} = 1$, Eq. (1) becomes

$$E_i = \frac{\hbar^3 c^4}{32 m^2 R^4} \sum_{\mathbf{k},\mathbf{q},\mathbf{q_o}} \frac{(\mathbf{q}.\mathbf{q_o})^2}{\omega_{\mathbf{q}}^2 \omega_{\mathbf{q_o}}^2} \times \frac{1}{2} \left| \sum_n e^{i\mathbf{q_o}.\mathbf{R_n}} \right|^2 \qquad (2)$$

$\sum_{\mathbf{k}} \mathbf{1} = 1$. Further,

$$\sum_{\mathbf{q}} \frac{(\mathbf{q}.\mathbf{q_o})^2}{\omega_{\mathbf{q}}^2 \omega_{\mathbf{q_o}}^2} = \left(\frac{2m}{\hbar}\right)^2 \frac{1}{q_o^2 c^3} \frac{\Omega}{(2\pi)^2} \int_0^{\pi} (\cos\alpha)^2 \sin\alpha d\alpha \int_0^{\infty} q dq = \left(\frac{2m}{\hbar}\right)^2 \frac{R^3}{9\pi c^3} \qquad (3)$$

For $\boldsymbol{R}_2 - \boldsymbol{R}_1 = \boldsymbol{R}$, we have

$$\sum_{\mathbf{q_o}} \left| \sum_n e^{i\mathbf{q_o}\mathbf{R_n}} \right|^2 = \sum_{\mathbf{q_o}} 2(1+\cos\Gamma) = 2\sum_{\mathbf{q_o}} (1+\cos(\mathbf{q_o}\mathbf{R})) =$$
$$2 + 2\frac{\Omega}{(2\pi)^2} \int_0^{0.94\pi/R} q_o^2 dq_o \int_0^{\pi} \cos(q_o R\cos\theta)\sin\theta d\theta = 3.3 \qquad (4)$$

where $\Gamma = \mathbf{q_o R_2} - \mathbf{q_o R_1}$. The interaction energy becomes

$$E_i = 0.00729 \frac{\hbar c}{R} = \alpha \frac{\hbar c}{R} \qquad \mathbf{(5)}$$

Taking the upper limit of $q_o$ as $0.94\pi/R$, which is with 6% lower than $\pi/R$, one obtains the value of $\alpha$ just as the experimental value. The relation (4) represents the Coulomb's law, which now is obtained without taking into account the electron charge concept. It was showed [10] that for charges of opposite sign the interaction energy (5) has the sign minus. In presence of a magnetic field in the above equation we introduce the potential vector and thus we substitute $\mathbf{q_o R}$ by $\mathbf{q_o R} - \frac{\mathbf{e}}{\hbar \mathbf{c}} \oint \mathbf{A} d\mathbf{l}$ and

$$\Gamma = \mathbf{q_o R_2} - \mathbf{q_o R_1} - \frac{e}{\hbar c}\left(\oint \mathbf{A}(\mathbf{R_2}) d\mathbf{l} - \oint \mathbf{A}(\mathbf{R_1}) d\mathbf{l}\right) \qquad (6)$$

We consider the potential vector $\mathbf{A} = (\boldsymbol{\mu} \times \mathbf{R})/R^3$ where $\boldsymbol{\mu}$ is the magnetic dipole moment and $\mathbf{R}$ is a vector from the middle of the loop to the observation point. The theory and experiment demonstrate that the free electron has a magnetic moment equal to the Bohr magneton $\mu_B$ and a spin momentum $\mathbf{s}$, the projections of which on a specified direction are $s_z = \pm\hbar/2 = \hbar m_s$ where $m_s = \pm 1/2$ is the spin quantum number. For $\mu_z^{(s)} = \mu_B g m_s$ with $g = 2$, one obtains

$$\Gamma = \frac{1}{2}(\mathbf{q}_2 + \mathbf{q}_1)(\mathbf{R}_2 - \mathbf{R}_1) + \frac{1}{2}(\mathbf{q}_2 - \mathbf{q}_1)(\mathbf{R}_2 + \mathbf{R}_1) - \frac{e}{\hbar c^2}\frac{e^2}{4\pi m}\frac{h}{e}\oint 2\frac{\mathbf{m}_{\mathbf{s2}} \times \mathbf{R}_{\mathbf{21}}}{R_{\mathbf{21}}^3} d\mathbf{l} - \frac{e}{\hbar c^2}\frac{e^2}{4\pi m}\frac{h}{e}\oint 2\frac{\mathbf{m}_{\mathbf{s1}} \times \mathbf{R}_{\mathbf{12}}}{R_{12}^3} d\mathbf{l} \tag{7}$$

where (h/e)$\mathbf{m}_{s2}$ and (h/e)$\mathbf{m}_{s1}$ are the flux vectors. For $\mathbf{q_1} = \mathbf{q_2} = \mathbf{q_o}$ resukts

$$\Gamma = q_o R\cos\theta - \Gamma_o$$
$$\Gamma_o = \frac{e^2}{mc^2}\left(\frac{4\pi m_{s2}}{R} - \frac{4\pi m_{s1}}{R}\right) \tag{8}$$

We have used the relation $\mathbf{q}_{\mathbf{o}}^{'} = \mathbf{q}_{\mathbf{o}} - \frac{e^2}{2mc^2}\frac{2m_s}{R^2}\mathbf{x}$ where **x** is a unit vector which is perpendicular to **R** and **μ**. The interaction enrgy between the two electrons when we take into account their magnetic moments is given by the expression

$$E_i = \frac{\hbar c}{244\pi R}(2 + 1.3\cos\Gamma_o) \tag{9}$$

where $\Gamma_o$ is given by Eq. (8). For $m_{s1} = m_{s2} = 1/2$ , one obtains $\Gamma_o = 0$, so that Eq. (9) reduces to Eq.(4), that is when the spins of the two electrons are orienred in the same direction there is not a modification of the Coukomb potential. When $m_{s1} = 1/2$, $m_{s2} = -1/2$ one obtains $\Gamma_o = 2\pi e^2/mc^2R$, so that for a certain value of R one obtains $\Gamma_o = \pi$, and the interaction energy between the two electrons is reduced by a factor of $0.7/3.3 \approx 1/5$.

However, like the electron the proton has a spin angular momentum with $s_p = 1/2$, and associated with this angular momentum is an intrinsic dipole moment

$$\vec{\mu}_{\mathbf{p}} = \frac{\gamma_p e}{mc}\mathbf{s}_{\mathbf{p}} \tag{10}$$

where M is the proton mass and $\gamma_p$ is a numerical factor known experimentally to be 2.7928. The magnetic moment of the electron moving around the proton is

$$\mu_e = \frac{e}{2mc}(\mathbf{L} + \mathbf{2S}) \tag{11}$$

where **L** is the orbital angular momentum and **S** is the epin angular momentum. For the hydrogen atom

$$\Gamma_o = \frac{e}{\hbar c}\oint\frac{(\vec{\mu}_\pi - \vec{\mu}_e)\times\mathbf{R}}{R^3} d\mathbf{l} = \frac{a}{r}$$
$$a = \frac{\pi e^2}{mc^2}\left(\frac{2m}{M}\gamma_p s_p - m_l - 2m_s\right) \tag{12}$$

$s_p=\pm 1/2$ is the proton spin quantum number, $m_s = \pm 1/2$ is the electron spin quanum number and $m_l$ is the magnetic quantum number of the electron. If in relation (11) we replace

$$\mathbf{L} + 2\mathbf{S} = g\mathbf{J} \qquad (11a)$$

where

$$g = 1 + \frac{s(s+1) + j(j+1) - l(l+1)}{2j(j+1)}$$

is the gyromagnetic factor, one obtains

$$a = \frac{\pi e^2}{mc^2}\left(\frac{2m}{M}\gamma_p s_p - g m_j\right) \qquad (12a)$$

where $m_j$ is the magnetic quantum number of the total angular momentum *j* of the electron. In Table I are given the values of parameter *a* for diferent states of the electron in hydrogen atom.

Table I

The values of the parameters *g* and *a* for the hydrogen atom energy states

| state | $m_l$ | $m_s$ | $s_p$ | g | *a*, $10^{-15}$m | |
|---|---|---|---|---|---|---|
| $nS_{1/2,1}$ | 0 | 1/ 2 | 1/ 2 | 2 | 8.8391067371418 | 8.839106731418 |
| $nS_{1/2,0}$ | 0 | 1/ 2 | -1/ 2 | 2 | 8.86601392707782 | 8.8660139270782 |
| $nP_{1/2}$ | 1 | -1/ 2 | 1/ 2 | 2/3 | 0.01345262138974 | 2.9373998490684 |
| $nP_{1/2}$ | 1 | -1/ 2 | -1/ 2 | 2/3 | 0.01345262138974 | 2.96430703390049 |
| $nP_{3/2,2}$ | 1 | 1/ 2 | 1/ 2 | 4/3 | 17.69166422937 | 17.691667069252 |
| $nP_{3/2,1}$ | 1 | 1/ 2 | -1/ 2 | 4/3 | 17.718569447215 | 17.718574259188 |
| $nD_{3/2,2}$ | 2 | -1/ 2 | 1/ 2 | 0.8 | 8.8391067371418 | 10.609618893964 |
| $nD_{3/2,1}$ | 2 | -1/ 2 | -1/ 2 | 0.8 | 8.86601392707782 | 10.6365259935 |
| $nD_{5/2,3}$ | 2 | 1/ 2 | 1/ 2 | 1.2 | 26.54422265475 | 26.544227401362 |
| $nD_{5/2,2}$ | 2 | 1/ 2 | -1/ 2 | 1.2 | 26.57112789753 | 26.571134591298 |
| $nF_{5/2,3}$ | 3 | -1/ 2 | 1/ 2 | 1.0285714285714 | 17.69166422937 | 22.750272973314 |
| $nF_{5/2,2}$ | 3 | -1/2 | -1/ 2 | 1.0285714285714 | 17.718569447215 | 22.77713016355 |
| $nF_{7/2,4}$ | 3 | 1/ 2 | 1/ 2 | 1.2380952380952 | 35.3967832921 | 38.347641177507 |
| $nF_{7/2,3}$ | 3 | 1/ 2 | -1/ 2 | 1.2380952380952 | 35.42368632291 | 38.374548367444 |

The 6$^{th}$ column is for the value of *a* given by Eq. (12) and the last column is for that given by Eq. (12*a*) in the case $m_j = j$.

## 3.THE ELECTRON ENERGY LEVELS IN THE HYDROGEN ATOM

For the radial wve function $\Psi = R(r)Y_l^m \exp(-iEt/\hbar)$, the non-relativistic Schrödinger equation for the hydrogrn atom becomes

$$\left[\frac{\hbar^2}{2m}\left(-\frac{d^2}{dr^2}-\frac{2}{r}\frac{d}{dr}+\frac{l(l+1)}{r^2}\right)-V_c(r)\right]R = ER \qquad (13)$$

Now we write R( r ) = $r^l\rho(r)$ where ρ(0) = 0. Eq. (13) is now

$$\left[-\frac{\hbar^2}{2m}\left(\frac{d^2}{dr^2}+\frac{2(l+1)}{r}\frac{d}{dr}\right)-\frac{2\hbar c}{144\pi r}\left(1+0.650132\cos(a/r)\right)\right]\rho = E\rho \qquad (14)$$

We are interested in the bound state solutions and therefore we assume $\rho(r) \sim e^{-\beta r}$ for $r\to\infty$, so that we try the solution ρ(r }=f(r )exp(-βr[1+0.65013266cos(a/r) +0.65013266sin(a/r)]), Eq (4) becomes

$$f'' - 2\beta[1+0.65013266\cos(a/r)+0.65013266\sin(a/r)]f' + \frac{1.30026632\beta a}{r}(-\sin(a/r)+\cos(a/r))f' +$$
$$0.65013266\frac{\beta a}{r^2}(\cos(a/r)-\sin(a/r))f + \beta^2[1+0.65013266\cos(a/r)+0.65013266\sin(a/r)]^2 f -$$
$$\frac{1.30026532\beta a}{r}(\cos(a/r)-\sin(a/r))[1+0.65013266\cos(a/r)+0.65013266\sin(a/r)]f +$$
$$0.65013266^2\beta^2\frac{a^2}{r^2}[\cos(a/r)-\sin(a/r)]^2 f - 0.65013266\frac{\beta a^2}{r^3}[-\cos(a/r)-\sin(a/r)]f +$$
$$\frac{2(l+1)}{r}f' - \frac{2(l+1)}{r}\beta[1+0.65013266\cos(a/r)+0.65013266\sin(a/r)]f +$$

$$1.30026532(l+1)\frac{\beta a}{r^2}\left[\cos(a/r)-\sin(a/r\right]f+\frac{2\hbar c}{144\pi r}\frac{2m}{\hbar^2}\left[1+0.65013266\cos(a/r)\right]f=-\frac{2m}{\hbar^2}Ef$$

(15)

To avoid f(r ) to diverge at infinity to overcome the wanted exponential supression, we require f(r ) to be a polynomial in r

$$f(r)=\sum_k c_k r^k \qquad (16)$$

The differential equaton then becomes

$$\sum_k \{c_k(k-1)r^{k-2}-2\beta\left[1+0.65013266\cos(a/r)+0.65013266\sin(a/r)\right]c_k kr^{k-1}+$$
$$1.30026532\beta a\left[\cos(a/r)-\sin(a/r)\right]c_k r^{k-2}+\beta^2[1+0.65013266\cos(a/r)+0.65013266\sin(a/r)]^2 c_k r^k-$$
$$1.30026532\beta^2 a\left[\cos(a/r)-\sin(a/r)\right]\left[1+0.65013266\cos(a/r)+0.65013266\sin(a/r)\right]c_k r^{k-1}+$$
$$(0.65013266)^2\beta^2 a^2\left[\cos(a/r)-\sin(a/r)\right]^2 c_k r^{k-2}-1.30026532\beta a^2\left[-\cos(a/r)-\sin(a/r)\right]c_k r^{k-3}+$$
$$2(l+1)c_k r^{k-2}-2(l+1)\beta\left[1+0.65013266\cos(a/r)+0.65013266\sin(a/r)\right]c_k r^{k-1}+$$
$$1.30026532(l+1)\beta a\left[\cos(a/r)-\sin(a/r)\right]c_k r^{k-2}+\frac{4mc}{144\pi\hbar}\left[1+0.65013266\cos(a/r)\right]c_k r^{k-1}+$$
$$\frac{2m}{\hbar^2}Ec_k r^k\}=0$$

(17)

At this stage we assume the constraint condition that the argument of *sine* and *cosine*, $a/r=a/n^2a_o$, where $n=l+k+1$ is the principal quantum number and $a_o$ is the Bohr radius. Collecting coefficients of $r^{k-1}$ the above equation gives us the recursion relation

$$c_{k+1}\{k(k+1)+1.30026532\beta a\left[\cos(a/n^2a_o)-\sin(a/n^2a_o)\right](k+1)+$$
$$0.65013266\beta a\left[\cos(a/n^2a_o)-\sin(a/n^2a_o)\right]+2(k+1)(l+1)+$$
$$0.65013266\beta^2a^2\left[\cos(a/n^2a_o)-\sin(a/n^2a_o)\right]^2+1.30026532(l+1)\beta a\left[\cos(a/n^2a_o)-\sin(a/n^2a_o)\right]\}-$$
$$c_k\{2\beta\left[1+0.65013266\cos(a/n^2a_o)+0.65013266\sin(a/n^2a_o)\right]k+$$
$$1.30026532\beta^2a\left[\cos(a/n^2a_o)-\sin(a/n^2a_o)\right]\left[1+0.65013266\cos(a/n^2a_o)+0.65013266\sin(a/n^2a_o)\right]+$$
$$2(l+1)\beta\left[1+0.65013266\cos(a/n^2a_o)+0.65013266\sin(a/n^2a_o)\right]-$$
$$\frac{4mc}{144\pi\hbar}\left[1+0.65013266\cos(a/n^2a_o)+0.65013266\sin(a/n^2a_o\right]\}+$$
$$c_{k-1}\{\beta^2\left[1+0.65013266\cos(a/n^2a_o)+0.65013266\sin(a/n^2a_o)\right]+\frac{2m}{\hbar^2}E\}-$$
$$c_{k+2}\{1.30024532\beta a^2\left[-\cos(a/n^2a_o)-\sin(a/n^2a_o)\right]\}=0 \tag{18}$$

We assume $c_{k+1}$-0, $c_{k+2}$=0, and

$$2\beta n+1.30026532\beta^2a\left[\cos(a/n^2a_o)-\sin(a/n^2a_o)\right]-$$
$$\frac{4mc}{144\pi\hbar}\frac{1+0.6513266\cos(a/n^2a_o)}{1+0.65013266\cos(a/n^2a_o)+0.65013266\sin(a/n^2a_o)}=0 \tag{19}$$
$$E=-\frac{\hbar^2}{2m}\beta^2\left[1+0.65013266\cos(a/n^2a_o)+0.65013266\sin(a/n^2a_o)\right]^2$$

whence

$$\beta=\frac{\sqrt{n^2+\frac{4mca}{144\pi\hbar}\times1.30026532\left[\cos(a/n^2a_o)-\sin(a/n^2a_o)\right]N}-n}{1..30026532a\left[\cos(a/n^2a_o)-\sin(a/n^2a_o)\right]} \tag{20}$$
$$N=\frac{1+0.65013266\cos(a/n^2a_o)}{1+0.65013266\cos(a/n^2a_o)+0.65013266\sin(a/n^2a_o)}$$

Further,

$$E=-\frac{h^2}{2m}\left[1+0.65013266\cos(a/n^2a_o)+0.65013266\sin(a/n^2a_o)\right]^2\times$$
$$\frac{\left[\sqrt{n^2+\frac{4mca}{144\pi\hbar}\times1.30026532\left[\cos(a/n^2a_o)-\sin(a/n^2a_o)\right]N}-n\right]^2}{(1.30026532)^2a^2\left[\cos(a/n^2a_o)-\sin(a/n^2a_o)\right]^2} \tag{21}$$

The values of the parameter *a* are given in Table I. For *a*→0, one obtains the usual formula

$$E = -\frac{\alpha^2 mc^2}{2n^2} \qquad (22)$$

By using series expansions

$$\begin{aligned}(1+x)^{1/2} &= 1 + \frac{x}{2} - \frac{x^2}{8} + \ldots \\ \cos(x) &= 1 - \frac{x^2}{2!} + \frac{x^4}{4!} - \ldots \qquad (23) \\ \sin(x) &= x - \frac{x^3}{3!} + \ldots\end{aligned}$$

Eq.(21) reduces to

$$E = -\frac{\alpha^2 mc^2}{2n^2}\left[1 - \frac{1.30026532}{3.30026532}\frac{a}{n^3 a_o}\left(1 - \frac{a}{n^2 a_o}\right)\frac{1.65013266}{1.65013266 + 0.65013266\frac{a}{n^2 a_o}}\right]^2 \qquad (26)$$

where $\alpha = 2\times 1.65013266/144\pi$. For hydrogen-like atoms $\alpha$ is replaced by $\alpha Z$. In Table II are presented the values of the hydrogen energy levels, which a calculated by using Eq. (26).The values from the second column are calculated by using for *a* relation (12), while the values from the third column are calculated by using Eq. (12*a*) for $m_j = j$.

Table II

Theoretical values for the hydrogen energy levels

| State | $-E$, eV | $-E$, eV |
|---|---|---|
| $1S_{1/2,1}$ | 13.596644259086 | 13.596644298919 |
| $1S_{1/2,0}$ | 13.596638854037 | 13.596638853414 |
| $2P_{1/2,1}$ | 3.3996083257277 | 3.3995898246396 |
| $2P_{1/2,0}$ | 3.3996083257277 | 3.3995896537839 |
| $2S_{1/2,1}$ | 3.3995524824729 | 3.3995524824676 |
| $2S_{1/2,0}$ | 3.3995523122557 | 3.39955231222263 |
| $2P_{3/2,2}$ | 3.3994964759455 | 3.399496475934 |
| $2P_{3/2,1}$ | 3.3994963057496 | 3.399496305714 |
| $3P_{1/2,1}$ | 1.5109370602782 | 1.5109346238136 |
| $3P_{1/2,0}$ | 1.5109370602782 | 1.5109346013966 |
| $3S_{1/2,1}$ | 1.510863419155 | 1.5109297061781 |
| $3S_{1/2,0}$ | 1.5109296837614 | 1.5109296837013 |
| $3P_{3/2,2}$ | 1.5109223300492 | 1.5109223300521 |
| $3P_{3/2,1}$ | 1.5109223076329 | 1.5109223076331 |
| $3D_{3/2,2}$ | 1.510863419155 | 1.5109282309608 |
| $3D_{3/2,1}$ | 1.5109296837614 | 1.5109282085009 |
| $3D_{5/2,3}$ | 1.5109149543317 | 1.51091495543264 |
| $3D_{5/2,}$ | 1.5109149319167 | 1.5109149319085 |
| $4P_{1/2,1}$ | 0.8499021000561 | 0.8499015218678 |
| $4P_{1/2,0}$ | 0.8499021000361 | 0.8499015165472 |
| $4S_{1/2,1}$ | 0.8499003548712 | 0.8499003548702 |
| $4S_{1/2,0}$ | 0.8499003495472 | 0.8499003495496 |
| $4P_{3/2,2}$ | 0.8498986044214 | 0.8498986044178 |
| $4P_{3/2,1}$ | 0.8498985990976 | 0.849898690874 |
| $4D_{3/2,2}$ | 0.8499003548712 | 0.8499000097754 |
| $4D_{3/2,1}$ | 0.8499003435472 | 0.8498999994549 |
| $4D_{5/2,3}$ | 0.8498970496022 | 0.8498968540181 |
| $4D_{5/2,2}$ | 0.8498970496022 | 0.8498968486979 |
| $4F_{5/2,3}$ | 0.8498986044214 | 0.8498976041829 |
| $4F_{5/2,2}$ | 0.8498985990976 | 0.8498975988626 |
| $4F_{7/2,4}$ | 0.8498985990976 | 0.8498941854055 |
| $4F_{7/2,3}$ | 0.8498950983502 | 0.8498945149142 |

We have used the following values of the constants: m=9.109389×$10^{-31}$ kg, c=2.997925×$10^{8}$m/s, ħ=1.054572×$10^{-34}$Js, $a_o$ = 0.529177×$10^{-10}$m. With these values of the constants one obtains $E_1 = \alpha^2 mc^2/2$ = 13.598433643441 eV. The obtained results are in a good agreement with experimental data. In Fig. 1 are presented some low –energy states of yhe hydrogen atom including fine structure, hyperfine structure and the Lamb shift. For the specific case of the ground state of the hydrogen atom (*n* =1) the energy separation between the states $1S_{1/2,0}$ and $1S_{1/2,1}$ is 5.6×$10^{-6}$ eV. The photon corresponding to the transitions between these states has wavelength close to 21 cm. This is the source of the famous "21 cm line" which is extremely useful to radio astronomers for tracking in the interstellar medium of galaxies. The separation between $2P_{3/2}$ and $2P_{1/2}$ states is $10^{-4}$ eV in the second column and the separation is 9.35×$10^{-5}$ eV in the third column, and is generated by the spin-orbit coupling. This appears to be two times larger than the experimental value. Lamb shift appears also as a natural result in our model. The difference in energy between the two energy levels $2S_{1/2}$ and $2P_{1/2}$ is 5.6×$10^{-5}$ eV in the second column and 3.7×$10^{-5}$ eV in the third column, and are some larger than the experimental value. In Table III we present the values of *a* and *E* when we use relation (12*a*) for $m < j$.

Table III

Values of the parameter *a* and of the energy *E* for $m < j$ in hydrogen atom

| State | $m_j$ | $a$,$10^{-15}$ m | -$E$, eV |
|---|---|---|---|
| $2P_{3/2,2}$ | 1/ 2 | 5.888253291095 | 3.3995711528654 |
| $2P_{3/2,}$ | 1/ 2 | 5.9151604830371 | 3.3995711528654 |
| $3P_{3/2,2}$ | 1/ 2 | 5.888253291695 | 1.5109321649756 |
| $3P_{3/2,1}$ | 1/ 2 | 5.9151604830371 | 1.510932142585 |
| $4P_{3/2,2}$ | 1/ 2 | 5.888253291095 | 0.84996093661 |
| $4P_{3/2,1}$ | 1/ 2 | 5.9151604830371 | 0.849900930455 |
| $3D_{3/2,2}$ | 1/ 2 | 3.5275705378802 | 1.5109341320458 |
| $3D_{3/2,1}$ | 1/ 2 | 3.5546777278078 | 1.5109341094463 |
| $4D_{3/2,2}$ | 1/ 2 | 3.5275705378803 | 0.8499014051670 |
| $4D_{3/2,1}$ | 1/ 2 | 3.5546777278078 | 0.8499013998068 |
| $3D_{5/2,3}$ | 3/2 | 15.921155007834 | 1.5109357104893 |
| $3D_{5/2,3}$ | 1/ 2 | 5.2980826043022 | 1.5109366300179 |
| $3D_{5/2,2}$ | 3/2 | 15.948062192762 | 1.5109237828262 |
| $3D_{5/2,2}$ | 1/ 2 | 5.3248897942298 | 1.510932634421 |
| $4D_{5/2,3}$ | 3/ 2 | 15.921155007834 | 0.8498989542875 |
| $4D_{5/2,3}$ | 1/ 2 | 5.2980826043022 | 0.8499010590711 |
| $4D_{5/2,2}$ | 3/2 | 15.948062192762 | 0.8498989491837 |
| $4D_{5/2,2}$ | 1/ 2 | 5.3248897942298 | 0.8498994046181 |
| $4F_{5/2,3}$ | 3/2 | 13.644782348006 | 0.8498994046181 |
| $4F_{5/2,3}$ | 1/ 2 | 4.5392917186926 | 0.8498012051090 |
| $4F_{5/2,2}$ | 3/2 | 13.671689535933 | 0.8498993992976 |
| $4F_{5/2,2}$ | 1/ 2 | 4.5661999086202 | 0.8499011997882 |
| $4F_{7/2,4}$ | 5/2 | 27.387328385376 | 0.8498966873162 |
| $4F_{7/2,4}$ | 3/2 | 16.42701559324 | 0.8498988544792 |
| $4F_{7/2,4}$ | 1/ 2 | 5.4667028011041 | 0.8499010217233 |
| $4F_{7/2,3}$ | 5/2 | 27.4142355575303 | 0.8498966819961 |
| $4F_{7/2,3}$ | 3/2 | 16.453922783168 | 0.8498988481588 |
| $4F_{7/2,3}$ | 1/ 2 | 5.493609990317 | 0.8499010164025 |

It is observed that the separation between $2P_{3/2}$ ($m_j$ = 1/ 2) and $2P_{1/2}$ states is $1.87\times10^{-5}$ eV which is by 2.5 times lower than the experimental value. It is possible that the levels $2P_{3/2}$ for $m_j$ = 3/2 and 1/ 2, respectively, participate to the transitions with a weight so that the result is that experimentally observed.

## 4. CONCLUSIONS

We have presented a thoery which includes in a simple formula fine and hyperfine structure, as well as the Lamb shift for the hydrogen atom. The theory is based on the modification of the Coulomb potential due to the interaction between the magnetic moments of the electron and proton, respectively. Every level associated with a particular set of quantum numbers *n, l* and *j* is split into two levels of slightly different energy depending on the relative orientation of the proton magnetic dipole with the electron state. The obtained results are in a good agreement with experimental data. For example, the separation energy between the two states of the ground state is close to the famous wavelength of a photon of 21 cm. The energy of the states $nP_{1/2}$ are lower than the energy of the states $nS_{1/2}$ because in the first case the interaction between the magnetic moments of the proton and the electron spins is diminished by the spin-orbit interaction. Some values of the separation between the energy states in our theory are overevaluated with respect to experimental data. This means that our theory may be improved.

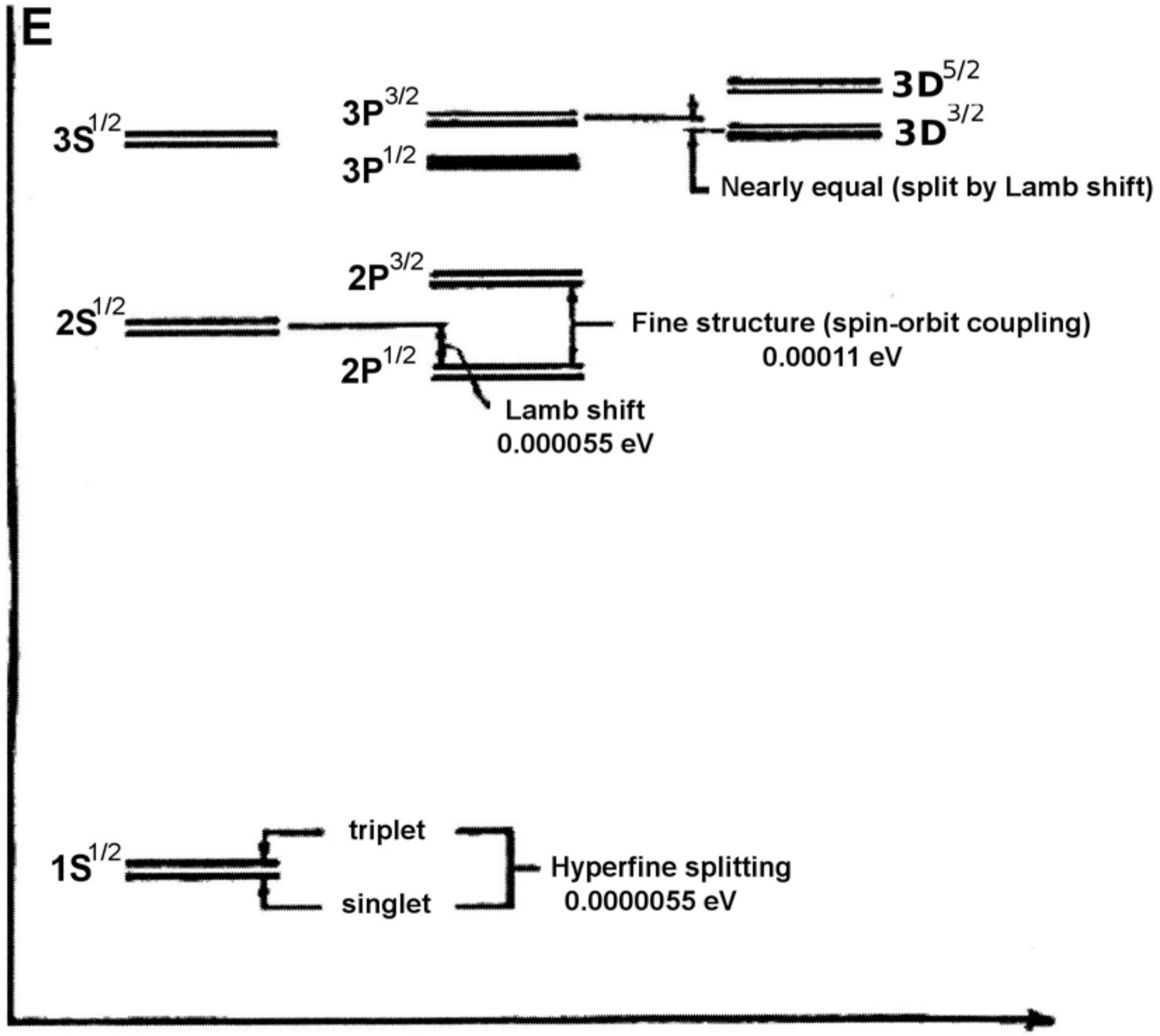

Fig. 1 Some low-energy states of the hydrogen atom, including fine structure, hyperfine structure, and the Lamb shift.